\documentclass[runningheads]{llncs}
\usepackage{graphicx}
\usepackage{latexsym}
\newcommand{\mysubsection}[1]{\vspace{0.3em} \noindent\textbf{#1}}
\usepackage{booktabs}
\usepackage{multicol}
\usepackage{floatrow}

\newfloatcommand{capbtabbox}{table}[][\FBwidth]

\usepackage{graphicx}
\usepackage{tikz}
\usepackage{pgfplots}
 \usetikzlibrary{spy}

    \newcommand{\Getxycoords}[3]{%
        \pgfplotsextra{%

            \pgfplotspointgetcoordinates{(#1)}%

             \global\pgfkeysgetvalue{/data point/x}{#2}%
             \global\pgfkeysgetvalue{/data point/y}{#3}%
         }%
    }
\usepackage{pgfplotstable}

\pgfplotsset{compat=1.16}

\begin{document}

\title{Chord Embeddings: Analyzing What They Capture and Their Role for Next Chord Prediction and Artist Attribute Prediction}

\titlerunning{Chord Embeddings: Analysis and Applications}

\authorrunning{A. Lahnala et al.}

\author{Allison Lahnala\inst{1} 
\and Gauri Kambhatla\inst{1}
\and Jiajun Peng\inst{1}
\and Matthew Whitehead\inst{2}
\and Gillian Minnehan\inst{1}
\and Eric Guldan\inst{1}
\and Jonathan K. Kummerfeld\inst{1}
\and Anıl Çamcı\inst{3}
\and Rada Mihalcea\inst{1}
}

\authorrunning{A. Lahnala et al.} 

\institute{Department of Computer Science \& Engineering \\
\and
School of Information\\
\and
Department of Performing Arts Technology\\
University of Michigan, Ann Arbor MI 48109, USA\\
\email{\{alcllahn,gkambhat,pjiajun,mwwhite,gminn,\\eguldan,jkummerf,acamci,mihalcea\}@umich.edu}}

\maketitle              

\begin{abstract}
Natural language processing methods have been applied in a variety of music studies, drawing the connection between music and language.
In this paper, we expand those approaches by 
investigating \textit{chord embeddings}, which we apply in two case studies to address two key questions: (1) what musical information do chord embeddings capture?; and (2) how might musical applications benefit from them?
In our analysis, we show that they capture similarities between chords that adhere to important relationships described in music theory. In the first case study, we demonstrate that using chord embeddings in a next chord prediction task yields predictions that more closely match those by experienced musicians. In the second case study, we show the potential benefits of using the representations in tasks related to musical stylometrics.

\end{abstract}

\section{Introduction}

Natural language processing (NLP) methods such as classification, parsing, or generation models have been used in many studies on music, drawing the connection that music is often argued to be a form of language.
However, while word embeddings 
are an important piece of almost all modern NLP applications, embeddings over musical notations have not been extensively explored.
In this paper, we explore the use of \textit{chord embeddings} and argue that it is yet another NLP methodology that can benefit the analysis of music as a form of language. Our objectives are (1) to probe embeddings to understand what musical information they capture, and (2) to demonstrate the value of embeddings in two example applications.

Using \texttt{word2vec}~\cite{mikolov2013efficient} to create embeddings over chord progressions, we first perform qualitative analyses of chord similarities captured by the embeddings using Principal Component Analysis (PCA).
We then present two case studies on chord embedding applications, first in next-chord prediction, and second in artist attribute prediction.
To show their value, we compare models that use chord embeddings with ones using other forms of chord representations.

In the next chord prediction study, 
we provide a short chord sequence and ask what the next chord should be.
We collected human annotations for the task as a point of reference.
By comparing model predictions with the human annotations, 
we observe that models using chord embeddings yield chords that are more similar to the predictions of more experienced musicians. 
We also measure the system's performance on a larger set drawn from real songs.
This task demonstrates a use case for chord embeddings that involves human perception, interaction, and composition.
For the artist attribute prediction study, we perform binary classification tasks on artist type (solo performer or group), artist gender (when applicable), and primary country of the artist.
Results on these tasks demonstrate that chord embeddings could be used in studies of musical style variations, including numerous studies in musicology.

This paper contributes analyses of the musical semantics captured in \textit{chord2vec} embeddings and of their benefits to two different computational music applications. We find that the embeddings encode musical relationships that are important in music theory such as the \textit{circle-of-fifths} and relative major and minor chords. The case studies provide insight into how musical applications may benefit from using chord embeddings in addition to NLP methods that have previously been employed.

\section{Related Work}

Methods for learning word embeddings \cite{mikolov2013efficient,pennington2014glove,peters-etal-2018-deep,devlin-etal-2019-bert} have been useful for domains outside of language (e.g., network analysis~\cite{grover2016node2vec}). 
Recent work has explored embeddings for chords, including an adaptation of \textit{word2vec}~\cite{madjiheurem2016chord2vec}, their use in a chord progression prediction module of a music generation application~\cite{brunner2017jambot}, and for aiding analysis and visualization of musical concepts in Bach chorales~\cite{phon2019exploring}.
However, understanding the musical information captured as latent features in the embeddings has been limited by the decision to ground evaluation in language modeling metrics (e.g., perplexity) rather than analyses of their behavior in downstream tasks. 
In this work, our first case study shows that language models with no remarkable differences in performance by perplexity exhibit remarkable relationships in their predictions to the experience of musicians, and furthermore we provide insights into what is captured by the embeddings.

NLP methods have benefited computational musicology topics such as authorship attribution~\cite{Stamatatos2009}, lyric analysis~\cite{fell2014lyrics}, and music classification tasks using audio and lyrics~\cite{mayer2011musical}.
In the task of composer identification, many approaches draw inspiration from NLP, applying musical stylometry features and melodic n-grams~\cite{brinkman2016musical,wolkowicz2008n,hillewaere2009melodic}. 
One study used language modeling methods on musical n-grams to perform composer recognition~\cite{hontanilla2013modeling} and another used a neural network over encodings of the pitches of musical pieces~\cite{kaliakatsos2010musical}.
Similar methods have been used to study stylistic characteristics of eight jazz composers~\cite{ogihara2008n,absolu2010analysis} over chord sequences similar to ours. 
While these studies operated on small datasets (on the order of hundreds of samples) to identify and analyze music of a small set of musicians, we use a large dataset (on the order of tens of thousands) and predict attributes of artists based on the music.

Our attribute prediction tasks are related to NLP work in authorship attribution and are motivated by studies on the connection between language and music in psychology~\cite{patel2003language,jancke2012relationship}, and the intersection of society, music, and language, or sociomusicology~\cite{feld1994music,shepherd1982theoretical}.
For instance, Sergeant \& Himonides studied the perception of gender in musical composition, and found no significant match between the listener's guess of a composer's gender and their actual gender \cite{sergeant2016genderedperception}.

\section{Dataset}\label{sec:dataset}
We compile a dataset of 92,000 crowdsourced chord charts with lyrics from the Ultimate Guitar website.\footnote{\url{https://www.ultimate-guitar.com/}}
We identify and remove duplicate songs\footnote{Sometimes multiple users submit chord charts for a song.} in our data using Jaccard similarity, then extract the chord progressions from our data representation to learn chord embeddings.

\begin{figure}
\centering
\begin{tikzpicture}
    \begin{scope}[
        spy using outlines={
            rectangle,
            magnification=2,
            connect spies,
            height=.8cm,
            width=3.3cm,
            blue,
        },
    ]
\begin{axis}[
    title={},
    xlabel={Chord song frequency rank},
    ylabel={Song frequency},
    ymin=0, ymax=60000,
    legend pos=north east,
    legend style={nodes={scale=0.7, transform shape}},
    legend image post style={mark=*},
    ymajorgrids=true,
    grid style=dashed,
    width=7cm,
    height=4.5cm
]
\addplot[domain=1:62, red, thin,] {184864 * x^-1.25};
\addplot[
    color=blue,
    mark=square,
    mark size=.2pt
    ]
    coordinates {
    (1,55910)(2,47202)(3,44696)(4,35516)(5,31462)(6,30808)(7,29401)(8,28065)(9,18827)(10,15891)(11,14969)(12,13207)(13,9336)(14,9218)(15,8251)(16,7153)(17,6283)(18,5975)(19,5908)(20,5751)(21,5362)(22,5016)(23,4778)(24,4730)(25,4348)(26,4325)(27,4013)(28,3948)(29,3841)(30,3513)(31,3167)(32,3070)(33,2761)(34,2755)(35,2644)(36,2229)(37,1978)(38,1805)(39,1805)(40,1676)(41,1652)(42,1462)(43,1422)(44,1366)(45,1361)(46,1216)(47,1201)(48,1171)(49,1148)(50,1138)(51,1138)(52,1121)(53,1089)(54,1070)(55,1069)(56,1048)(57,998)(58,982)(59,915)(60,900)(61,892)
    };
    \node[label={180:{\scalebox{.5}{G}}}, inner sep=0pt] at (axis cs:1,55910) {};
    \node[label={180:{\scalebox{.5}{C}}},inner sep=0pt] at (axis cs:2,47202) {};
    \node[label={180:{\scalebox{.5}{D}}},inner sep=0pt] at (axis cs:3,44696) {};
    \node[label={180:{\scalebox{.5}{Am}}},inner sep=0pt] at (axis cs:5,31462) {};
    \node[label={190:{\scalebox{.5}{F}}},inner sep=0pt] at (axis cs:6,30808) {};
    \node[label={205:{\scalebox{.5}{Em}}},inner sep=0pt] at (axis cs:7,29401) {};
    \node[label={0:{\scalebox{.5}{Bm}}},inner sep=0pt] at (axis cs:9,18827) {};
     \node[label={0:{\scalebox{.5}{Dm}}},inner sep=0pt] at (axis cs:10,15891) {};  
    \node[label={90:{\scalebox{.3}{Fm}}},inner sep=-1pt] at (axis cs:25,4348) {}; 
    \node[label={90:{\scalebox{.3}{Ab}}},inner sep=-1pt] at (axis cs:34,2755) {}; 
    \node[label={90:{\scalebox{.3}{Bbm}}},inner sep=-1pt] at (axis cs:40,1676) {}; 
    \node[label={90:{\scalebox{.3}{Db}}},inner sep=-1pt] at (axis cs:44,1366) {}; 
    
    \legend{$184864x^{-1.25} R^2 = .936$}

    \coordinate (point) at (axis cs:34.5,5000); 
    \Getxycoords{point}{\PointX}{\PointY}
    \coordinate (spy point) at (axis cs:34.5,30000);
    \spy on (point) in node (spy) at (spy point);

\end{axis}
\end{scope}

\end{tikzpicture}
    \caption{Number of chords in the dataset and a fitted trendline with the parameters given in the figure for the 61 most common chords, showing a power law distribution.
    A sample of chords are labeled that also appear in the embedding visualization in Figure~\ref{tab:PCA_plots}.}
    \label{fig:chord_dfs}
\end{figure}
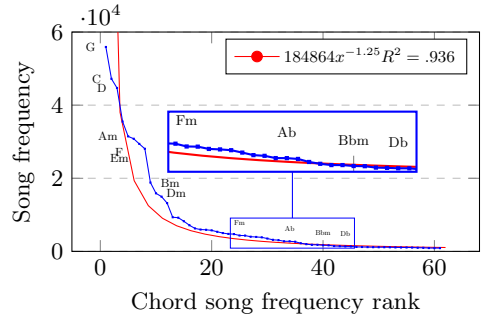

We remove songs with fewer than six chords, leaving us with a final set of 88,874 songs and 4,913 unique chords. 
The chords' song frequencies are distributed according to a power law distribution, much like Zipf's word frequency law exhibited in natural language. 
Figure~\ref{fig:chord_dfs} shows the song frequency and the power law trend line fitted to the top 61 chords for demonstration, though the trend is even stronger over all chords.
Beyond there existing many possible chord notations, we observe a relation between chord frequency and physical difficulty of playing the chord on the guitar or other instruments (e.g., \texttt{G}, \texttt{C}, and \texttt{D}), and variations in notation.

\section{Chord Embeddings}\label{sec:chord_embeddings}

Word embeddings have been used extensively to represent the meaning of words based on their use in a large collection of documents. 
We consider a similar process for chords, forming representations based on their use in a large collection of songs. 

We create chord embeddings for chords that appear in at least 0.1\% of our songs (237 chords) using the continuous bag of words model (\texttt{CE}$_{cbow}$) and the \textit{skip-gram} language model (\texttt{CE}$_{sglm}$) from \texttt{word2vec}~\cite{mikolov2013efficient}, a widely used method for creating word embeddings.
For \texttt{CE}$_{cbow}$, a target chord is predicted based on context chords, while \texttt{CE}$_{sglm}$ is the reverse: context chords are predicted given a target chord.
In both cases, this has the effect of learning representations that are more similar for chords that appear in similar contexts.
We tested context sizes of two, five, and seven, and varied the vector dimensions between 50, 100, 200, and 300, but observed only minor differences across different models, and chose to use a context window of five and a vector dimension of 200.

\begin{figure*}[t]
\centering
\scalebox{0.9}{
\begingroup
\setlength{\tabcolsep}{6pt} 
\renewcommand{\arraystretch}{0}
    \begin{tabular}{cc}
         \includegraphics[width=0.48\linewidth]{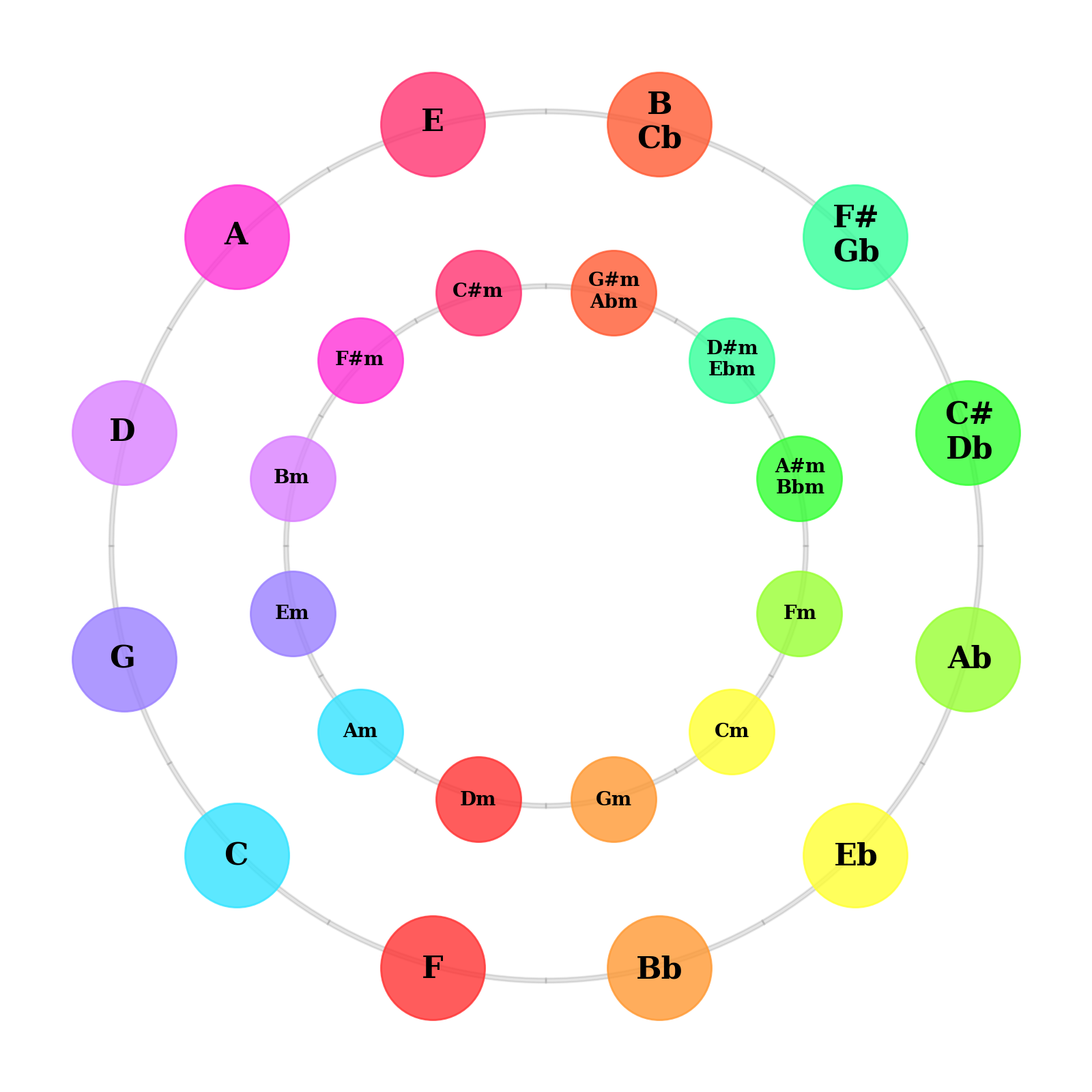} &
         \includegraphics[width=0.48\linewidth]{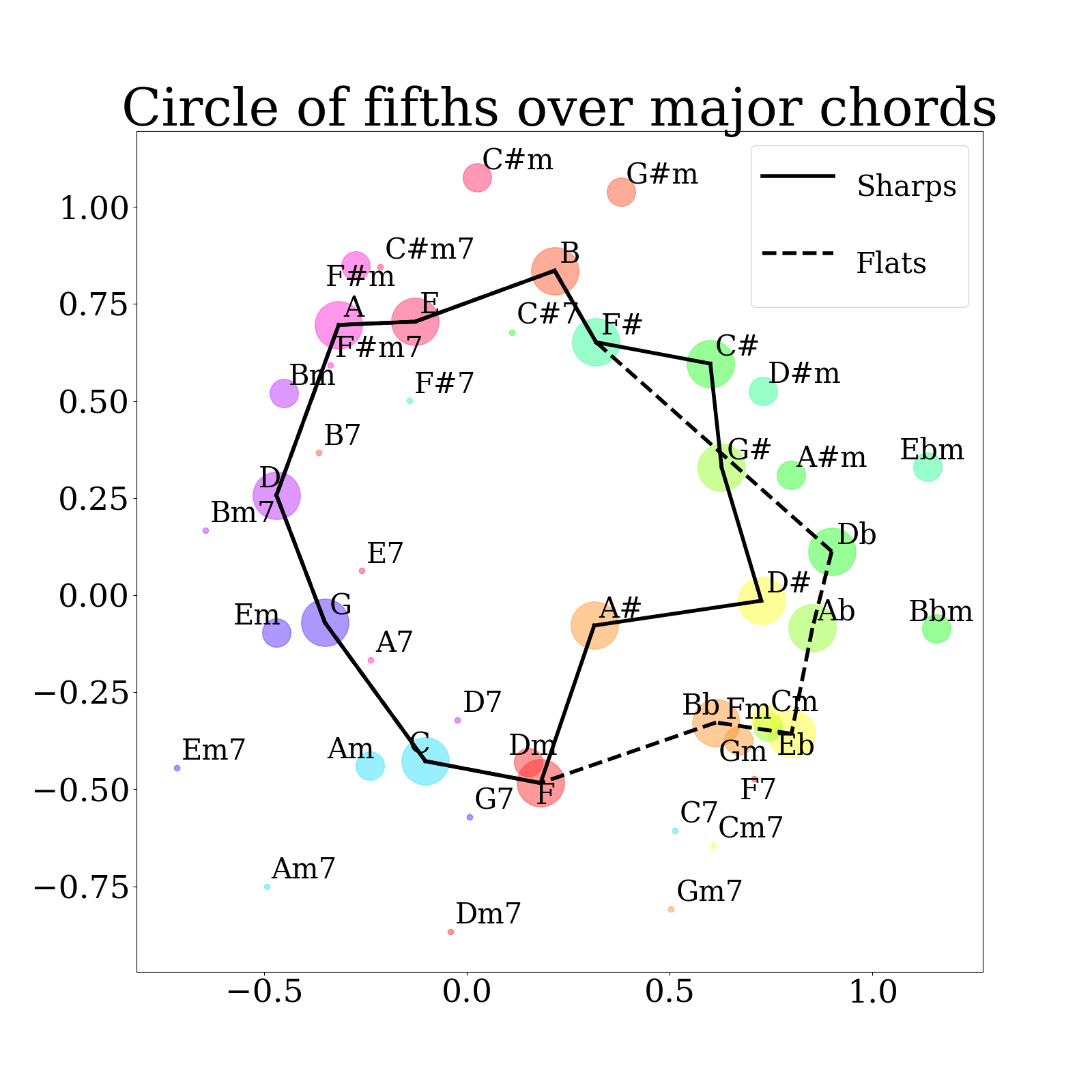} \\
         (a) & (b) \\
         \includegraphics[width=0.48\linewidth]{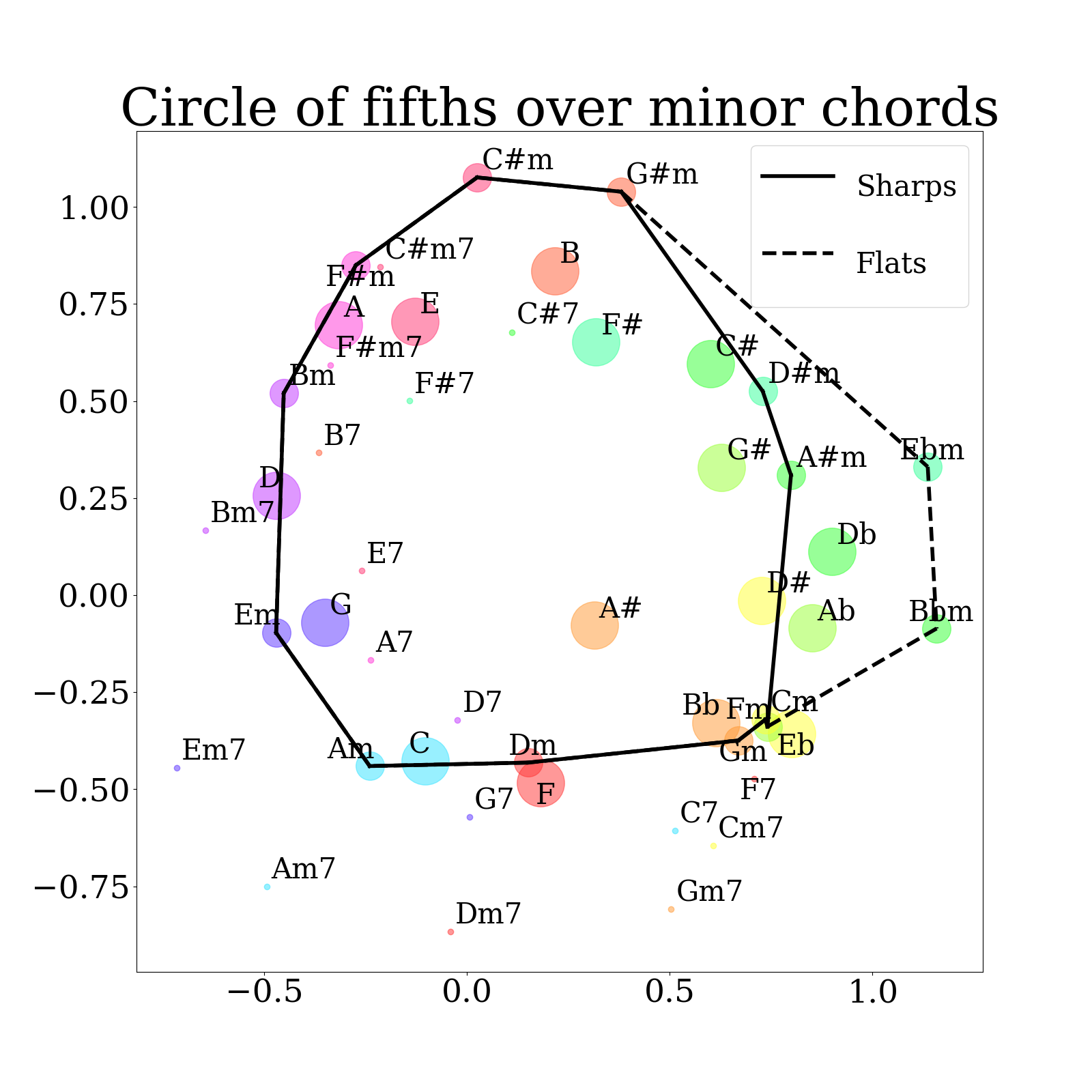} &
         \includegraphics[width=0.48\linewidth]{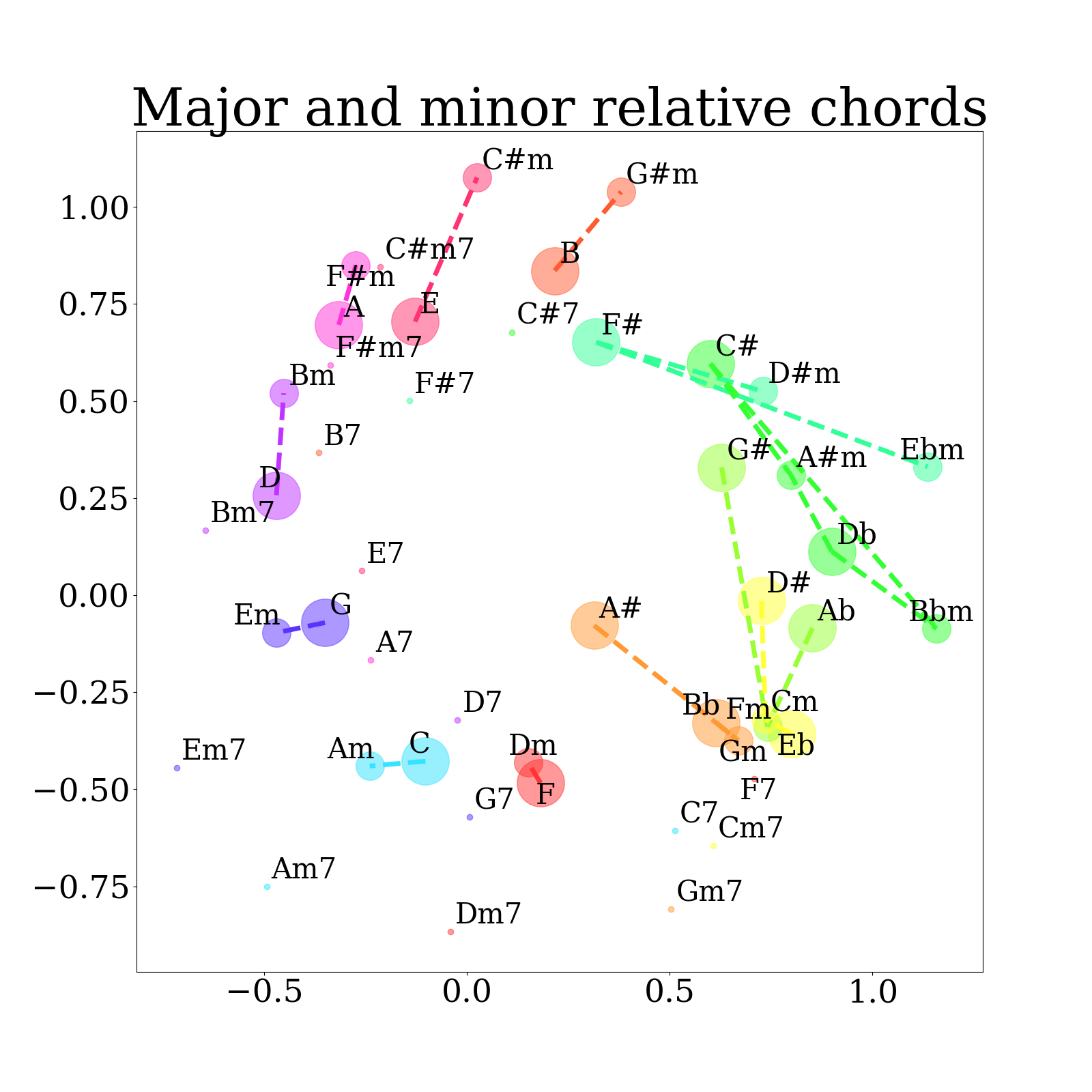}  \\
         (c) & (d) 
    \end{tabular}
    \endgroup
    \caption{In (a), we show the circle of fifths with the same colors as in (b), (c) and (d), which show the same 2-dimensional PCA projection of the chord embedding space with lines denoting the circle of fifths over major chords (b) and minor chords (c), and lines denoting major-minor relatives (d).
    }
    }
    \label{tab:PCA_plots}
\end{figure*}

\subsection{Qualitative Analyses} \label{sec:pca_discussion}
To better understand the information encoded in chord embeddings, we perform a qualitative analysis using PCA and present a 2D projection for the \texttt{CE}$_{sglm}$ model in Figure~\ref{tab:PCA_plots} (our main observations are consistent for the case of \texttt{CE}$_{cbow}$). 

We observe that chords that form a fifth interval are closer together, which suggests that the embeddings capture an important concept known as the \textit{circle of fifths}. Fifth-interval relationships serve broad purposes in tonal music (music that adheres to a model of relationships to a central tone). Saker~\cite{saker1992theory} encapsulates their structural importance by stating that ``circle of fifths relationships dominate all structural levels of tonal compositions'' and that ``the strongest, most copious harmonic progressions to be found in tonal music are fifth related.''
The \textit{circle of fifths} relationship is observed in our chord embeddings over different chord \textit{qualities},\footnote{\textit{Qualities} refers to sound properties that are consistent across chords with different roots but equidistant constituent pitches. The interaction of intervals between pitches determines the quality.} specifically over major chords (highlighted in Figure~\ref{tab:PCA_plots}b), minor chords (highlighted in Figure~\ref{tab:PCA_plots}c), major-minor 7 chords, and minor 7 chords.
For both chord qualities, the layout captured by the chord embeddings is similar to the ideal/theoretical circle of fifths, illustrated in Figure~\ref{tab:PCA_plots}a.
This pattern is particularly interesting as it does not follow the style of word analogy patterns observed in language.
This makes sense, as the ``is-a-fifth" relation forms a circle in chords, whereas word analogies connect pairs of words without forming a circle.

Additionally, we observe that relative major and minor chords\footnote{\textit{Relative} refers to the relation between the chords' roots, in which the scale beginning on the minor chord's root shares the same notes as the scale beginning on the major chord's root, but the ordering of the notes give different qualities to the scales.} appear relatively close together in the embedding space, as shown by their proximity in the PCA plots (highlighted in Figure~\ref{tab:PCA_plots}d).
We also observe that enharmonics, notes with different names but the same pitch, are often close together.
Not only that, but there is a consistent pattern in the positioning of enharmonics, with sharps to the left and flats to the right.

These observations suggest that chord embeddings are capable of representing musical relationships that are important to music theory.
Transitions between the tonic (I), dominant (V), and subdominant (IV) chords of a scale are prescriptive components in musical cadences~\cite{randel1999harvard}.
Since these chords frequently appear in the same context, their embeddings are more similar.
A common \textit{deceptive cadence} is a transition between the fifth and sixth root chords of a scale~\cite{owen2000music}. An example progression with a deceptive cadence in C major is \texttt{C}-\texttt{F}-\texttt{G}-\texttt{Am}; these chords appear in a similar neighborhood in the PCA plots.
Because these chords frequently co-occur in music, the embeddings capture a relationship between them.

We also note that relationships for chords that are used more frequently are more strongly represented.
The major and minor relative pairs (\texttt{G}, \texttt{Em}), (\texttt{C}, \texttt{Am}), and (\texttt{D}, \texttt{Bm}), are among the top ten chords ranked by song frequency (Figure~\ref{fig:chord_dfs}) and have clear relations in Figure~\ref{tab:PCA_plots}.
In contrast, the pairs (\texttt{Ab}, \texttt{Fm}) and (\texttt{Db}, \texttt{Bbm}) are ranked lower and their minor-major relative relationship appears weaker by their distance.

\subsection{Alternative Representations}\label{sec:alt_representations} 
In addition to chord embeddings, we also explore two other chord representations: Pitch Representations (\texttt{PR}) and Bag-of-Chords (\texttt{BOC}).
For a fair comparison, we use the same vocabulary of 237 chords for these representations.

\mysubsection{Pitch Representations.}\label{sec:CNN_PR} A chord's identity is determined by its pitches, so we test if the individual pitches provide a better representation of a chord as a whole than our chord embeddings. This method represents each chord by encoding each of its pitches by their index in the chromatic scale $\{\texttt{C} = 1, \texttt{C\#} = 2, \cdots, \texttt{B} = 12\}$. 
The pitches are in order of the triad, followed by an additional pitch if marked, and by one extra dimension for special cases.\footnote{Special cases include: the ``*'' marking on a chord, which is a special marker specific to the ultimate-guitar.com site; ``UNK'' which we use to replace chords that do not meet the 0.1\% document frequency threshold; and ``H'' and ``Hm'' which indicates ``hammer-ons'' in the notation on ultimate-guitar.com} 
Additional pitches are indicated by the interval relative to the root of the pitch that is being added. We also represent chord inversions, e.g., the chord \texttt{G/B} which is an inversion of \texttt{G} such that the bottom pitch is \texttt{B}. 

\mysubsection{Bag-of-Chords.} In this representation, each chord is represented as a ``one-hot'' vector, where the vectors have length equal to the vocabulary size.
We consider two ways of determining the value for a chord.
For \texttt{BOC}$_{count}$, we use the frequency of a chord in a song, divided by the number of chords in the song.
For \texttt{BOC}$_{tfidf}$, we use the TF-IDF of each chord (term-frequency, inverse document frequency).

\section{Case Study One: Next Chord Prediction}
In this section, we present our first case study, which investigates if there is a relationship between chord embedding representations and the ways humans perceive and interact with chords. We test the use of chord representations for predicting the most likely next chord following a given sequence of chords, and then compare these to human-annotated responses.

We train a next chord prediction model using a \textit{long short-term memory} model~\cite{hochreiter1997long} (LSTM).\footnote{We use an open-source repository of neural language models \url{https://github.com/pytorch/examples/blob/master/word\_language\_model/model.py}}
We follow standard practice and do not freeze the embeddings, meaning the chord representations undergo updates throughout training, adjusting to capture the musical features most important for the task.
Our main model uses the pre-trained chord embeddings to initialize the chord prediction architecture. We test both \texttt{CE}$_{cbow}$ and \texttt{CE}$_{sglm}$ embeddings, and will refer to these models by these acronyms.
We also define a baseline model, where the encoder is randomly initialized (denoted \texttt{NI}, for no initialization).
Finally, we also evaluate a model where we initialize the encoder with the pitch representations introduced in Section~\ref{sec:alt_representations} (denoted \texttt{PR}). 

We divide our data into three sets, with 69,985 songs (80\%) for training, 8,748 songs (10\%) for validation, and 8,748 (10\%) for testing.
We train using a single GPU with parameters: epochs = $40$, sequence length = $35$, batch size = $20$, dropout rate = $0.2$, hidden layers = $2$, learning rate = $20$, and gradient clipping = $0.25$.

\subsection{Human Annotations}

To evaluate the next chord prediction models, we collect data with a human annotation task in which annotators are asked to add a new chord at the end of a chord progression.
For example, given the progression ``\texttt{A}, \texttt{D}, \texttt{E}," they must pick a chord that would immediately follow \texttt{E}.
They are also asked to pick one or two alternatives to this selection.
Continuing the example, if an annotator provides \texttt{E7} and \texttt{A}, then the chord progressions they have specified are ``\texttt{A}, \texttt{D}, \texttt{E}, \texttt{E7},'' and ``\texttt{A}, \texttt{D}, \texttt{E}, \texttt{A}.''
The annotators are given a tool to play 48 different chords (all major, minor, major-minor 7, and minor 7 chords) so they can hear how different options would sound as part of the progression.\footnote{We did not limit our next chord prediction models to these 48 chords.} They were given a total of 39 samples shown in the same order to all annotators. The samples were chosen randomly from our entire dataset of songs, permitting only one sequence to come from a single song, and requiring they contain the same 48 chords. We presented sequences of length three and six as we expect that patterns in the given sequence affect the responses.

\mysubsection{Participants.} The annotators were first asked to estimate their expertise in music theory on a scale from $0$ - $100$, 
where $0$ indicates no knowledge of music theory, $25$ - $75$ indicates some level of knowledge from pre-university training though self-teaching, private lessons/tutoring, or classroom settings, and $75$ - $100$ indicates substantial expertise gained by formal university studies, performing and/or composing experience. 
They were given the option to provide comments about how they estimated their expertise.
We collected this information because we expected that the annotations provided by a participant may vary depending on their background education in music theory.
It also allows us to perform comparisons of our system with sub-groups defined by self-reported knowledge.
Nine participants provided complete responses, 
with expertise ratings of 
0, 0, 10, 10, 19, 25, 25, 50 and 73.
For the following analyses, we define a \textit{beginner} set containing annotators who provided 0 for their self-rating, an \textit{intermediate} set containing annotators with ratings $> 0$ and $< 50$, and an \textit{expert} set containing annotators whose ratings are at least 50. 

\mysubsection{Inter-annotator Agreement.} 
For pairwise agreement, we compute the proportion of chord progressions in which a pair of annotators provided the same chord, averaged over all pairs of annotators.
The pairwise agreement across all annotators is 22.51, it is 23.08 for the beginner set, 25.38 for the intermediate set, and 17.95 for the expert set.

To account for responses of similar but not identical chords (discussed in Section~\ref{sec:chord_pred_eval}), we measure pairwise agreement on response pitches. We compute the fraction of matching pitches for a pair of annotators' responses for a given chord progression, averaged over all pairs of annotators.
The pairwise pitch agreement score for all annotators is 38.00, it is 37.01 for the beginner set, 40.90 for the intermediate set, and 33.59 for the expert set.
The average number of unique chords used by each annotator is 30.2, it is 35.5 for the beginner set, 27.4 for intermediate set, and 32.0 for the expert set.

\subsection{Evaluation Metrics}\label{sec:chord_pred_eval}

The main objective of this case study is to investigate whether chord similarities captured by our embeddings reflect human-perceived similarities.
We use the chord-prediction systems to perform the same task given to the annotators. Each model provides a probability distribution over the full set of chords, therefore we treat the chords with highest probabilities as each model's selection.

We evaluate the predictions with the following metrics, which are inspired by the metrics employed by the Lexical Substitution task~\cite{mccarthy2009english} but modified for our setup, which weight more frequent responses higher:

$Match_{best}$:
For each example we calculate the fraction of people who included the model's top prediction in their answer.
These values are then averaged over all examples.

$Match_{oo4}$:
This adds together values for the previous metric across the model's top four predictions.

$Mode_{best}$:
The fraction of cases in which the top model prediction is the same as the most common annotator response, when there is a single most common response. 

$Mode_{oo4}$:
The fraction of cases in which one of the top four model predictions is the same as the most common annotator response.

Note that only 25 out of the 39 examples had a unique most common response.
Of these, 20 had a chord chosen by three or four annotators, and five had a chord chosen by five to seven annotators.
The rest of the examples are not considered in the $Mode$ metrics.

\textit{Pitch Matches}:
The metrics above penalise all differences in predictions equally, even though some chord pairs are more different than others, e.g., \texttt{A7} and \texttt{A} differ only in the addition of a pitch, whereas \texttt{B} and \texttt{A} share no pitches.
To address this, we use a metric that is the total number of pitches that match for each question between the model's top response and the annotator's first response.
We calculate this separately per-annotator, and then average across annotators ($PM_{ave}$). 

$Loss$ and $Perplexity (PPL)$:
These are two measures from the language modeling literature that we apply to see how well the models do on the true content of songs.
Note that this evaluation is over a different set: 8,748 randomly chosen songs that are not included in model training.

\begin{table}
    \centering
    \begin{tabular}{ccc}
        \begin{tabular}{lrrrrr}
            \toprule
              & \multicolumn{2}{c}{$Match$} & \multicolumn{2}{c}{$Mode$} & $PM$ \\
              & $best$ & $oo4$ & $best$ & $oo4$ & $ave$ \\
            \midrule
            \multicolumn{6}{c}{All} \\
            \midrule
            \texttt{NI} & 7.52 & 30.10 & \textbf{32.00} & 64.00 & 48.33 \\
            \texttt{PR} & 7.49 & 29.64 & 24.00 & \textbf{72.00} & \textbf{51.11}\\
            \texttt{CE$_{cbow}$} & 7.45 & 29.82 & 16.00 & 64.00 & 47.78 \\
            \texttt{CE$_{sglm}$} & \textbf{7.60} & \textbf{30.22} & 12.00 & 68.00 & 49.00 \\
            \midrule
            \midrule
            \multicolumn{6}{c}{Intermediate} \\
            \midrule
            \texttt{NI} & 8.17 & 32.68 & \textbf{33.33} & \textbf{71.43} & 48.80 \\
            \texttt{PR} & 8.19 & 32.42 & 23.81 & \textbf{71.43} & \textbf{53.60} \\
            \texttt{CE$_{cbow}$} & 8.04 & 32.14 & 14.29 & 66.67 & 47.20 \\
            \texttt{CE$_{sglm}$} & \textbf{8.34} & \textbf{33.19} & 14.29 & \textbf{71.43} & 50.00 \\
            \bottomrule
        \end{tabular} & & \begin{tabular}{lrrrrr}
            \toprule
              & \multicolumn{2}{c}{$Match$} & \multicolumn{2}{c}{$Mode$} & $PM$ \\
              & $best$ & $oo4$ & $best$ & $oo4$ & $ave$ \\
            \midrule
            \multicolumn{6}{c}{Beginner} \\
            \midrule
            \texttt{NI} & \textbf{5.61} & \textbf{22.44} & 0.00 & \textbf{44.44} & 40.50 \\
            \texttt{PR} & \textbf{5.61} & \textbf{22.44} & \textbf{11.11} & 33.33 & \textbf{43.00}\\
            \texttt{CE$_{cbow}$} & 4.97 & 19.87 & 0.00 & 22.22 & 39.50 \\
            \texttt{CE$_{sglm}$} & 5.29 & 21.15 & 0.00 & 33.33 & 42.00 \\
            \midrule
            \midrule
            \multicolumn{6}{c}{Expert} \\
            \midrule
            \texttt{NI} & 7.92 & 31.67 & 28.57 & \textbf{85.71} & 55.00 \\
            \texttt{PR} & 7.72 & 30.26 & 28.57 & \textbf{85.71} & 53.00 \\
            \texttt{CE$_{cbow}$} & \textbf{8.56} & \textbf{34.23} & \textbf{42.86} & \textbf{85.71} & \textbf{57.50} \\
            \texttt{CE$_{sglm}$} & 8.15 & 32.18 & 14.29 & 57.14 & 53.50 \\
            \bottomrule
        \end{tabular} \\
    \end{tabular}
    \caption{Results for all models when compared with different sets of annotators based on their expertise.}
    \label{tab:lexsub_expertise_groups}
\end{table}

\subsection{Results}

\mysubsection{Match and Mode.}
Table~\ref{tab:lexsub_expertise_groups} reports metrics for each system when compared with the beginner set, intermediate set, expert set, and full set of annotators.
When evaluating against all annotators, \texttt{CE}$_{sglm}$ is best in $Match_{best}$ and $Match_{oo4}$, \texttt{NI} is best in $Mode_{best}$, and \texttt{PR} is best in $Mode_{oo4}$.

In the results for subsets of annotators, all systems tend to match experts better than the beginner or intermediate groups.
In particular, \texttt{CE}$_{cbow}$ obtains the lowest scores when evaluated against the beginner group, but the highest when evaluated against the expert group.
To investigate this pattern, we compute $Match_{best}$ and $Match_{oo4}$ for each individual annotator and then the Spearman Correlation Coefficient $r_s$ between these metrics and their expertise.
In Table~\ref{tab:spearman}, we observe strong significant correlation between the \texttt{CE}$_{cbow}$ and expertise, and no significant correlation for the other models. This can be explained by the fact that the models are trained on a large collection of songs composed by experts, and the chord embeddings seem to capture the chord use style of the experts.

\mysubsection{Pitch Matches.}
We report $PM_{ave}$ for the models and expertise groups in Table~\ref{tab:lexsub_expertise_groups}.
Similarly to the $Match$ and $Mode$ metrics, we observe that all models perform better when compared to annotators with higher expertise, and the differences between groups is most extreme with \texttt{CE$_{cbow}$}.
Table~\ref{tab:spearman} shows the results of correlation analysis for pitch matches (Pearson Correlation Coefficient $r_p$), with a significant linear correlation for only \texttt{CE}$_{cbow}$.
This trend is visually shown in Figure~\ref{fig:lin_reg_pitch_matches}. 

\begin{figure}
\begin{floatrow}
\ffigbox{%
\begin{tikzpicture}
    \begin{axis}[
        xlabel={Annotator expertise},
        ylabel={Pitch matches},
        legend pos=north west,
        legend style={nodes={scale=0.7, transform shape}},
        legend image post style={mark=*},
        ymajorgrids=true,
        grid style=dashed,
        width=5cm,
        height=4cm
    ]
    \addplot[domain=0:75, red,thick] {0.27788406351635736 * x + 41.2320642816147};
    \addplot[
        color=blue,
        only marks,
        ]
        coordinates {
        (0,38)(0,41)(10,47)(10,43)(19,51)(25,46)(25,49)(50,53)(73,62)
        };
        \legend{$.3x + 41$}
    \end{axis}
    \end{tikzpicture}

}{%
  \caption{Total pitch matches between annotator and output of \texttt{CE}$_{cbow}$ model, plotted by the annotator's expertise. The red line is the line of best fit computed by linear regression.}%
  \label{fig:lin_reg_pitch_matches}
}
\capbtabbox{%
\begingroup
\setlength{\tabcolsep}{1pt}
\renewcommand{\arraystretch}{1}
\begin{tabular}{lcccccc}
\toprule

& $best \: r_s$ & p-val & $oo4 \: r_s$ & p-val & $PM \: r_p$  & p-val \\
\midrule
\texttt{NI} & 0.44 & 0.24 & 0.44 & 0.24 & 0.71	& 0.03 \\
\texttt{PR} & 0.64 & 0.06 & 0.57 & 0.11 & 0.43 & 0.25 \\
\texttt{CE}$_{cbow}$ & \textbf{0.72} & \textbf{0.03} & \textbf{0.72} & \textbf{0.03} & \textbf{0.94} & \textbf{2e-4} \\
\texttt{CE}$_{sglm}$ & 0.41 & 0.28 & 0.41 & 0.28 & 0.42 & 0.27  \\
\bottomrule
&&&&&& \\
&&&&&& \\
\end{tabular}
\endgroup
}{%
  \caption{Correlation coefficients for expertise and Match and Pitch Match metrics.}%
  \label{tab:spearman}
}
\end{floatrow}
\end{figure}

\begin{table}
    \centering
    \begin{tabular}{l c c}
    \toprule
     & Test Loss & Test PPL \\
     \midrule
\texttt{NI} & 1.42 & 4.16 \\
\texttt{PR} & 1.44 & 4.20 \\ 
\texttt{CE}$_{cbow}$ & 1.44 & 4.22 \\
\texttt{CE}$_{sglm}$ & 1.42 & 4.15 \\
\bottomrule
    \end{tabular}
    \caption{Loss and perplexity metrics for the chord prediction models on a held-out test set. }
    \label{tab:lm_test_results}
\end{table}

\mysubsection{Loss and Perplexity.}
Table~\ref{tab:lm_test_results} shows the results on our 8,748 song test set.
All models perform similarly in this setting. 

\subsection{Discussion}

We observe that automatic models produce predictions that resemble human responses. Comparing against annotators grouped by expertise, \texttt{CE}$_{cbow}$ compares best to the high expertise group for all metrics, while the best model varies among the other groups. \texttt{CE}$_{cbow}$'s predictions also correlate significantly with pitch matches and annotator expertise.
\texttt{NI} and \texttt{PR} achieve the highest $Mode_{best}$ and $Mode_{oo4}$ scores, 
however, fewer samples are considered because only twenty-five had a unique $mode$ chord across the annotators and only five of these samples had more than half the annotators agree. Additionally, the chord symbol based metrics are strict, requiring an exact match on chords, and had lower interannotator agreement than the pitch-based metrics.

While \texttt{CE}$_{cbow}$'s predictions exhibit a strong pitch-match correlation, \texttt{CE}$_{sglm}$'s predictions exhibit no significant correlate at all. However, differences between the \texttt{CE}$_{cbow}$ and \texttt{CE}$_{sglm}$ embeddings may not be as apparent in other downstream applications; in fact, by the perplexity and test loss metrics shown in Table~\ref{tab:lm_test_results}, there is barely a difference between these two, or any, models. Investigating key differences between these embedding models in musical contexts is a direction for future work.

\section{Case Study Two: Artist Attribute Prediction}

Our second case study introduces the task of performing artist attribute prediction, demonstrating that these chord representations could be used more broadly in tasks involving musical stylometry and musical author profiling. 
With binary classifiers using our chord representations, we predict three attributes as separate tasks: gender (male or female), performing country (U.S.~or U.K.), and type of artist (group or solo artist).

\mysubsection{Data.} 
For these experiments, we augment the dataset with information obtained with the MusicBrainz API,\footnote{\url{https://musicbrainz.org/}} which includes the song artist's location, gender, lifespan, tags that convey musical genres, and other available information for 35,351 English songs (identified using Google's Language Detection project~\cite{nakatani2010langdetect}). From this extracted information, we choose artist type, performing country, and gender because of the sufficient quantity of data available with these attributes enabling the tasks; we note that tasks dedicated to genre or time period are of interest for future investigations, and our preliminary experiments using the artists' lifespan and tags as proxies for time period and genre indicated these tasks are promising use cases for chord embeddings.

We use the top two most frequent classes of each attribute, and balance the data to have the same number of examples for each class.
For artist type, there are 20,000 songs per class (group and solo).
For performing country, there are 8,000 songs per class (U.S.~and U.K.).
For gender, there are 6,000 songs per class (male and female).
The number of samples varies because of differences in the raw class counts and because not all songs have a label for each property.

\mysubsection{Experimental Setup.} We build two binary classifiers and compare their performance with each chord representation. The first uses logistic regression (LR) over a single vector for each song by aggregating chord representations. We also experimented with an SVM classifier, but LR was more efficient with minimal performance trade-offs. The \texttt{BOC} methods are defined in Section~\ref{sec:alt_representations}. The \texttt{PR} method aggregates the chords with a many-hot encoding vector counting each chord pitch, normalized by the total number of pitches. The \texttt{CE} methods aggregate chord embeddings by max-pooling, taking the most extreme absolute value in each dimension across all chords. 

The second classifier is a Convolutional Neural Network (CNN) that considers the chords in sequences. We experimented with an LSTM, and found that the CNN functions better for these tasks. 
We use the CNN model for sentence classification by Kim~\cite{kim2014convolutional}\footnote{CNN model is built on \url{https://github.com/Shawn1993/cnn-text-classification-pytorch}} over the chord progressions for each song, using the same \texttt{NI}, \texttt{PR}, \texttt{CE}$_{sglm}$, and \texttt{CE}$_{cbow}$ representations from the first case study. 

For our model parameters, chosen in preliminary experiments on a subset of the data, we use L2 regularization for the LR classifier, and the CNN model uses filter window sizes 3, 4, 5 with 30 feature maps, drop-out rate 0.5 and Adam~\cite{kingma2014adam} optimization. The sequence limit is 60 chords, cutting off extra chords and padding when there are fewer. 

\begin{table}
    \centering
    \begin{tabular}{l  r r r}
    \toprule
  Model  &  Gender & Country & Artist Type\\
  \midrule
  \multicolumn{4}{c}{\textit{Logistic regression}} \\
  \midrule
    \texttt{BOC}$_{count}$ &  \textsuperscript{*$\dagger\ddagger$}\textbf{57.53} & \textsuperscript{$\dagger$}55.71 & \textsuperscript{*$\dagger$}\textbf{57.04} \\
    \texttt{BOC}$_{tfidf}$  &  \textsuperscript{$\dagger$}55.94 & \textsuperscript{$\dagger$}55.17 & \textsuperscript{$\dagger$}56.06 \\
    \midrule
    \texttt{PR}  &  52.32 & 53.13 & 53.75 \\
    \midrule
    \texttt{CE}$_{cbow}$  &  \textsuperscript{*$\dagger$}56.90 & \textsuperscript{$\dagger$}55.37 & \textsuperscript{*$\dagger$}56.92 \\
    \texttt{CE}$_{sglm}$ &  \textsuperscript{$\dagger$}56.37  & \textsuperscript{$\dagger$}\textbf{55.85} & \textsuperscript{*$\dagger$}56.88 \\
    \bottomrule
\end{tabular}
    \hspace{5mm}
    \begin{tabular}{l  r r r}
    \toprule
  Model  &  Gender & Country & Artist Type\\
  \midrule
    \multicolumn{4}{c}{\textit{CNN}} \\
    \midrule
    \texttt{NI} & \textsuperscript{$\dagger$}58.93  & \textsuperscript{$\dagger$}56.79 & \textsuperscript{$\dagger$}59.20  \\
     &  &  &   \\
     \midrule
    \texttt{PR}  & 57.98 & 55.31 & 57.54 \\
    \midrule
    \texttt{CE}$_{cbow}$  &  58.67 & \textsuperscript{$\dagger$}57.30 & \textsuperscript{$\dagger$}58.92 \\
    \texttt{CE}$_{sglm}$  &  \textsuperscript{$\dagger$}\textbf{58.95} & \textsuperscript{$\dagger$}\textbf{57.54}  & \textsuperscript{$\dagger$}\textbf{59.29} \\
    \bottomrule
\end{tabular}
    \vspace{1mm}
    
{\footnotesize*$p<0.05$ over \texttt{BOC}$_{tfidf}$, $\dagger p<0.05$ over \texttt{PR}, $\ddagger p<0.05$ over \texttt{CE}$_{sglm}$, $\mathsection p<0.05$ over \texttt{NI}}
    \caption{Accuracy scores from 10-fold cross validation in artist gender, country, and type prediction tasks. The significance tests are performed among the logistic regression models and CNN models separately.}
    \label{tab:Gender}
\end{table}

\subsection{Results and Analyses}

Table~\ref{tab:Gender} shows the models' accuracy scores from experiments using 10-fold cross validation. \texttt{CE}$_{sglm}$ CNN is the top performer for all tasks, significantly outperforming CNN \texttt{PR} and all LR models for all tasks, and \texttt{NI} for country.\footnote{By a paired t-test with p $< .05$.} 
All models outperform a random baseline (50\%) significantly in each task.
In each task, CNN models significantly outperform their LR counterparts.

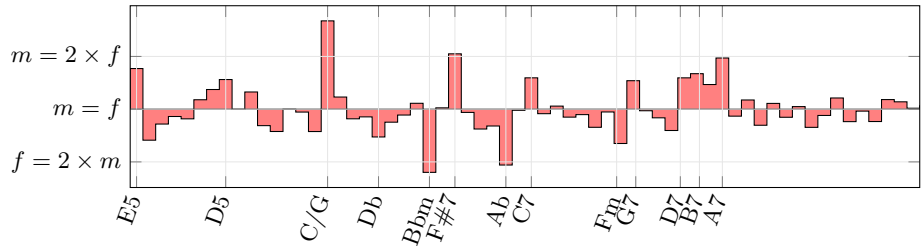
\begin{figure}
    \centering
\begin{tikzpicture}
	\begin{axis}[
	    width=.99\linewidth,
        height=4cm,
		const plot,
		area style,
		grid=both,
        grid style={line width=.1pt, draw=gray!20},
		enlarge x limits=false, 
		xticklabels={E5, D5, C/G, Db, Bbm, F\#7, Ab, C7, Fm, G7, D7, B7, A7},xtick={0.5, 7.5, 15.5, 19.5, 23.5, 25.5, 29.5, 31.5, 38.2, 39.7, 43.2, 44.7, 46.5},
        x tick label style={rotate=70, anchor=east},
        yticklabels={$f=2\times m$, $m=f$,$m=2\times f$},ytick={-1,0,1},
        y tick label style={anchor=east, },]
    ]
	\addplot[color=black, fill=red, fill opacity=0.5] coordinates
		{(0,0.770281457)(1,-0.5884449324)(2,-0.2823324092)(3,-0.1391218369)(4,-0.1835040785)(5,0.1784560694)(6,0.3726490066)(7,0.561732969)(8,-0.002876196525)(9,0.3257492423)(10,-0.312483336)(11,-0.4236208434)(12,-0.002311385053)(13,-0.05519415959)(14,-0.4245723943)(15,1.674305475)(16,0.2299133637)(17,-0.1828276614)(18,-0.1463440616)(19,-0.5272977153)(20,-0.2468591692)(21,-0.1114625528)(22,0.1113433591)(23,-1.19837565)(24,0.02496688742)(25,1.049298384)(26,-0.06152864981)(27,-0.3764604936)(28,-0.3193690681)(29,-1.059441118)(30,-0.02139718961)(31,0.594123069)(32,-0.08571428571)(33,0.05788414787)(34,-0.1512489244)(35,-0.1034563246)(36,-0.3439620955)(37,-0.05322726665)(38,-0.6511686723)(39,0.5397746002)(40,-0.03006244045)(41,-0.1642078731)(42,-0.4047265254)(43,0.5939604202)(44,0.6728448152)(45,0.467039849)(46,0.9701547207)(47,-0.132537352)(48,0.174102942)(49,-0.3052385306)(50,0.1092605497)(51,-0.153679115)(52,0.04846625572)(53,-0.3449354585)(54,-0.1192839877)(55,0.2120932751)(56,-0.2365637747)(57,-0.03718840245)(58,-0.2346961809)(59,0.1817797062)(60,0.1393538188)(61,0.01797896783)(62,0.06883565098)} 
		\closedcycle;
	\end{axis}
\end{tikzpicture}
    \caption{Chord variations by gender, ordered from least to greatest by song frequency. The labeled chords are at least 1.5 more salient for one gender than the other.}
    \label{fig:chord_variations}
\end{figure}

For insight into the models' performance, we analyze the gender prediction task, the only attribute where the LR \texttt{CE}$_{cbow}$ and LR \texttt{BOC}$_{count}$ predictions differed significantly. 
First, we compare the rate of use of each chord between genders. To show the differences in Figure~\ref{fig:chord_variations}a, we divide the higher rate by the lower rate, subtract one to set equal use to zero, and flip the sign when female use is higher.
We observe greater variations among chords with lower song frequency. For instance, \texttt{C/G}, \texttt{F\#7}, \texttt{Bbm}, and \texttt{Ab} are twice as salient for one gender than the other. 
The highest variation among the top $20$ chords reaches 1.5 times more salient, and for the top $10$, 1.2 times more salient for one gender. 

To investigate the impacts of the musical relationships captured in embeddings (Section~\ref{sec:chord_embeddings}) to the \texttt{CE} models, we also compared use of five \textit{chord qualities}.
Figure~\ref{fig:chord_variations}b shows higher relative frequency of suspended and diminished chords among the songs of male artists, augmented and minor chords among the female artist songs, and fairly similar use of major chords.

\begin{figure}
\centering
\begin{tikzpicture}
    \begin{axis}[
            width=7cm,
            height=4cm,
            grid=both,
            grid style={line width=.8pt, draw=gray!20},
            ybar=-1.0cm,
            bar width=1cm,
            symbolic x coords={Aug,Min,Maj,Sus,Dim},
            yticklabels={$f=1.25\times m$, $m=f$,$m=1.25\times f$},ytick={-.25,0,.25},
            y tick label style={anchor=east, }
        ]
        \addplot[red,fill,fill opacity=0.5,draw=black] coordinates{(Aug,-0.43118924995)};
        \addplot[red,fill,fill opacity=0.5,draw=black] coordinates{(Min,-0.24330941958)};
        \addplot[red,fill,fill opacity=0.5,draw=black] coordinates{(Maj,0.07870050177)};
        \addplot[red,fill,fill opacity=0.5,draw=black] coordinates{(Sus,0.184789286)};
        \addplot[red,fill,fill opacity=0.5,draw=black] coordinates{(Dim,0.3794521737)};
    \end{axis}
\end{tikzpicture}
\caption{Variation in chord quality usage by gender by ratio of use percentage.}
\label{fig:chord_gender_diffs}
\end{figure}
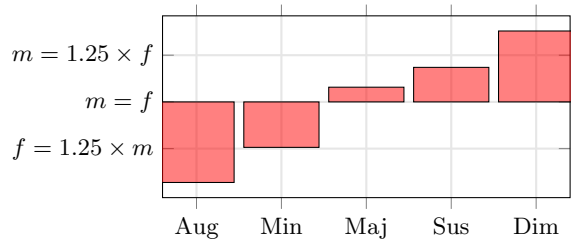
\subsection{Discussion}

To our knowledge, this is the first time that an association between chord representations
and author attributes has been explored. Each model for each attribute showed significant improvement over a random baseline of 50\%, indicating there are quantifiable differences in our music data between the genders, countries, and artist types. 
In addition to the tasks we presented, we also observed improvement when using the system to predict the life-period of the artist and their associated genres.
As life-period is a proxy for the music's time period, chord embeddings could benefit future work in musical historical analysis. 

For gender, the variation of rare chords may contribute to \texttt{BOC}$_{count}$ outperforming \texttt{CE}$_{cbow}$.
However, \texttt{CE}$_{cbow}$ significantly outperforms \texttt{BOC}$_{tfidf}$ which gives more weight to rare chords by their inverse-document frequency. This suggests that chord rarity is not the only critical feature. 
The variations of chord quality use may contribute to the performance of the \texttt{CE} models as the embeddings capture musical relationships.

LR \texttt{PR} consistently underperforms all other models, which may indicate the importance chord structures.
Different chords with the same pitches (e.g., \texttt{G} and \texttt{G/B}) have the same \texttt{PR} vector. Chords with overlapping pitches have similar \texttt{PR} vectors.
However, CNN \texttt{PR}, which performs closer to the others, encodes pitch orderings (Section~\ref{sec:CNN_PR}) and \texttt{BOC} methods encode chord symbols which indicate structure. \texttt{CE} representations are learned from chord symbols, likely capturing contextual functions of chord structures. These functions would matter for the CNN models which make predictions from chord sequences rather than a single aggregated vector. Since we observed the best performance by CNN \texttt{CE}$_{sglm}$, there suggests the importance of contextual semantics of chord structures.
A deeper study into structural semantics captured by chord embeddings is a direction for future work.

\section{Conclusion}

In this paper, we presented an analysis of the information captured by chord embeddings and explored how they can be applied in two case studies.
We found that chord embeddings capture chord similarities that are consistent with important musical relationships described in music theory.
Our case studies showed that the embeddings are beneficial when integrated in models for downstream computational music tasks.
Together, these results indicate that chord embeddings are another useful NLP tool for musical studies.
The code to train chord embeddings and the resulting embeddings, as well as the next-chord annotations are publicly available from \url{https://lit.eecs.umich.edu/downloads.html}.

\subsubsection{Acknowledgements.} We would like to thank the anonymous reviewers and the members of the Language and Information Technologies lab at Michigan for their helpful suggestions. We are grateful to MeiXing Dong and Charles Welch for helping with the design and interface of the next-chord annotation task. This material is based in part upon work supported by the Michigan Institute for Data Science, and by Girls Encoded and Google for sponsoring Jiajun Peng through the Explore Computer Science Research program. Any opinions, findings, and conclusions or recommendations expressed in this material are those of the authors and do not necessarily reflect the views of the Michigan Institute for Data Science, Girls Encoded or Google.

%
%
%
\bibliographystyle{splncs04}
\bibliography{references}

\end{document}